\documentclass{ws-procs961x669}            
\begin{document}
\def\apj{ApJ}  
\def\apjl{ApJ} 
\def\apss{Astrophys. Space Sci.} 
\def\prl{Phys.~Rev.~Lett.} 
\def\pr{Phys.~Rev.} 
\def\prc{Phys.~Rev.~C} 
\def\prd{Phys.~Rev.~D} 
\def\pre{Phys.~Rev.~E} 
\def\prx{Phys.~Rev.~X} 
\def\aap{A\&A}   
\def\physrep{Phys.~Rep.}   
\def\nphysa{Nucl.~Phys.~A}   
\def\nphysb{Nucl.~Phys.~B}   
\def\mnras{MNRAS}             
\def\epja{Eur.~Phys.~J.~A}
\def\cpc{Chin.~Phys.~C}
\def\jpgnpp{J.~Phys.~G:~Nucl.~Part.~Phys.} 
\def\jcap{J.~Cosmolol.~Astropart.~P.} 
\def\plb{Phys.~Lett.~B} 
\def\zap{Z. Astrophys.} 
\def\sovast{Sov. Astron.} 
\def\ssr{Space Sci. Rev.} 

\def\figsubcap#1{\par\noindent\centering\footnotesize(#1)}

\title{Electron Captures and Stability of White Dwarfs}

\author{N. Chamel$^*$, L. Perot}

\address{Institut d'Astronomie et d'Astrophysique, Universit\'e Libre de Bruxelles, \\ 
CP-226, 1050 Brussels, Belgium\\
$^*$E-mail: nicolas.chamel@ulb.be}

\author{A.~F. Fantina}
\address{Grand Acc\'el\'erateur National d'Ions Lourds (GANIL), CEA/DRF -
	CNRS/IN2P3, Boulevard Henri Becquerel, 14076 Caen, France}
\address{Institut d'Astronomie et d'Astrophysique, CP-226, Boulevard du Triomphe, Universit\'e Libre de Bruxelles, 1050 Brussels, Belgium}

\author{D. Chatterjee, S. Ghosh}
\address{Inter-University Centre for Astronomy and Astrophysics, Post Bag 4, Ganeshkhind, Pune University Campus, Pune, 411007, India}

\author{J. Novak, M. Oertel}
\address{LUTH, Observatoire de Paris, PSL Research University, CNRS, Universit\'e Paris Diderot, Sorbonne Paris Cit\'e, 5 place Jules Janssen, 92195 Meudon, France}

\begin{abstract}
Electron captures by atomic nuclei in dense matter are among the most important processes governing the late evolution of stars, limiting in particular the stability of white dwarfs. Despite considerable progress in the determination of the equation of state of dense Coulomb plasmas, the threshold electron Fermi energies are still generally estimated from the corresponding $Q$ values in vacuum. Moreover, most studies have focused on nonmagnetized matter. However, some white dwarfs are endowed with magnetic fields reaching $10^9$~G. Even more extreme magnetic fields might exist in super Chandrasekhar white dwarfs, the progenitors of overluminous type Ia supernovae like SN 2006gz and SN 2009dc. The roles of the dense stellar medium and magnetic fields on the onset of electron captures and on the structure of white dwarfs are briefly reviewed. New analytical formulas are derived to evaluate the threshold density for the onset of electron captures for arbitrary magnetic fields. Their influence on the structure of white dwarfs is illustrated by simple analytical formulas and numerical calculations. 
\end{abstract}

\keywords{White dwarf; Chandrasekhar limit; Electron capture; Magnetic field.}

\bodymatter

\section{Introduction}

In 1926, the British physicist Ralph Fowler showed that the energy and the pressure of matter in the dense core of a white dwarf remain finite at zero temperature due to quantum mechanics. Considering that electrons are no longer bound to nuclei under such extreme conditions, he derived the equation of state of a degenerate electron Fermi gas~\cite{fowler1926}. Another British physicist, Edmund Clifton Stoner, calculated the structure of white dwarfs with uniform density in 1929~\cite{stoner1929}. The same year, 
Wilhelm Anderson, a physicist at the University of Tartu in Estonia, first stressed the importance of taking into account the relativistic motion of electrons~\cite{anderson1929}. He showed that the mass of a white dwarf tends to some finite value as the electron concentration increases, namely $M_\text{And}\sim 0.69 M_\odot$ assuming an electron fraction $y_e=0.4$ (with $M_\odot$ the mass of the Sun). Improving Anderson's approximate treatment, Stoner~\cite{stoner1930} found $M_\text{St}\simeq 1.10 M_\odot$. 
Soon afterwards, the young Indian physicist Subrahmanyan Chandrasekhar solved the hydrostatic equilibrium equations for an  ultrarelativistic electron Fermi gas using the theory of polytropes~\cite{chandra1931}. His numerical result, $M_\text{Ch}\simeq 0.91 M_\odot$ (for $y_e=0.4$), thus differed by less than 20\% from that obtained earlier by Stoner, a remarkably close agreement as pointed out by Chandrasekhar himself. A year later, the Russian physicist Lev Landau showed that any degenerate star has a maximum mass, which can be expressed in terms of the fundamental constants as
\begin{equation}\label{eq:mass-limit-Landau}
	M_\text{L}=3.1 \frac{m_\text{P}^3}{m^2} y_e^2 \, ,
\end{equation}
where we have introduced the Planck mass
\begin{equation}
m_\text{P}=\sqrt{\frac{\hbar c}{G}} \, ,
\end{equation}
($\hbar$ is the Planck-Dirac constant, $c$ is the speed of light, $G$ is the constant of gravitation) and $m$ is the average mass per nucleon. 
Incidentally, such a scaling was already apparent in Stoner's analysis of a uniform density star. Combining his equations (17) and (19b) leads to 
\begin{equation}\label{eq:mass-limit-Stoner}
M_\text{St}=\frac{15\sqrt{5\pi}}{16}\frac{m_\text{P}^3}{m^2}y_e^2  \, .
\end{equation}
Landau however did not believe in the physical reality of this limit and even invoked some violation of the laws of quantum mechanics inside massive stars to prevent them from collapsing (see Ref.~\citenum{yakovlev2013} for a historical perspective on Landau's contribution). As early as 1928, the process by which atoms are crushed at high densities was actually discussed by the Russian physicist Yakov Frenkel~\cite{frenkel1928}. More importantly, he calculated the equation of state of an electron Fermi gas for an arbitrary degree of relativistic motion as well as the correction due to electrostatic interactions between electrons and atomic nuclei. He also studied the conditions for which incompressible ``superdense stars'' can exist and derived indirectly the mass limit. Correcting for an error in his expression for the gravitational energy\footnote{the factor $5/3$ in his key equation (19a) should read $3/5$; this error goes back to his incorrect expression (18) for the gravitational energy.} and neglecting the electrostatic correction lead to the same result as the one published by Stoner in 1930. 
Frenkel's pioneer work remained unnoticed during several decades, and is still not very well-known today~\cite{yakovlev1994}. 

The analyses of Anderson and Stoner showed that the maximum mass of white dwarfs is only reached asymptotically when the electron concentration tends to infinity and the stellar radius goes to zero. 
This conclusion was later confirmed by the numerical calculations of Chandrasekhar for more realistic density profiles~\cite{chandra1935}. However, soon after the discovery of the neutron by James Chadwick in 1932, it was realized that matter becomes predominantly composed of neutrons at high densities~\cite{sterne1933,hund1936}. In December 1933, during 
a meeting of the American Physical Society at Stanford, Wilhelm Baade and Fritz Zwicky predicted the existence of \emph{neutron stars} 
formed from the catastrophic gravitational collapse of stars during supernova explosions~\cite{bz1934}. Baade and Zwicky were apparently unaware of the studies about white dwarfs. 
The connection was first made by Landau~\cite{landau1938} and Gamow~\cite{gamow1939}. At a conference in Paris in 1939, Chandrasekhar also pointed out~\cite{chandra1941}: 
\begin{quote}
If the degenerate core attains sufficiently high densities, the protons and electrons will combine to form neutrons. This would cause 
a sudden diminution of pressure resulting in the collapse of the star to a neutron core. 
\end{quote}
In 1956, the French physicist Evry Schatzman~\cite{schatzman1956} showed that the central density of white dwarfs is limited by the onset of electron captures by nuclei, implying that the radius of the most massive white dwarfs remains \emph{finite}. Detailed calculations of the structure of white dwarfs taking into account matter neutronization were performed at the end of the 1950s and at the beginning of the 1960s~\cite{schatzman1958,harrison1958,hamada1961}. 
Electron captures are now known to play a key role in the late stages of stellar evolution (see Ref.~\citenum{langanke2021} for a recent review). 

Although the electrostatic correction to the equation of state of an electron Fermi gas was calculated long ago~\cite{frenkel1928} (see Ref.~\citenum{skye2021} and references therein for the latest developments on the equation of state of dense Coulomb plasmas), the threshold electron Fermi energy $\mu_e$ for the onset of electron captures by nuclei is still generally estimated from the corresponding $Q$ value in \emph{vacuum}. Moreover, the presence of magnetic fields is usually ignored. However, a significant fraction of white dwarfs have been found~\cite{ferrario2015} to have magnetic fields up to $10^9$~G, and potentially much stronger fields may exist in their core~\cite{fujisawa2012}. Moreover, it has been recently proposed that very massive so-called super-Chandrasekhar white dwarfs 
(with a mass $M\gtrsim 2M_\odot$) endowed with extremely strong magnetic fields could be the progenitors of overluminous type Ia 
supernovae like SN~2006gz and SN~2009dc~\cite{das2013,das2015,sathyawageeswar2015} (see also Refs.~\citenum{bera2014,franzon2015,bera2016,bera2017}). The existence of such stars was actually first studied much earlier by Shul'man~\cite{shulman1976}, who found that the maximum mass of degenerate stars could be increased by two orders of magnitude if the magnetic field is strongly quantizing~\cite{shulman1989,shulman1992}. However, the stability of such super Chandrasekhar white dwarfs will still be limited by electron captures~\cite{chamel2013,chamel2014}. Detailed calculations of the global structure of these stars taking these processes into account have been carried out in Refs.~\citenum{chatterjee2017,otoniel2019}. 

In this paper, we review our recent studies of the role of electron-ion interactions and magnetic fields in the onset of electron captures in cold white-dwarf cores~\cite{fantina2015,chatterjee2017}. We also present more accurate and more general formulas for the threshold density and pressure, applicable not only to nonmagnetic and strongly magnetized white dwarfs but also to stars with intermediate magnetic field strengths. The impact of electron captures on the global structure of white dwarfs is discussed and new numerical results are presented.

\section{Core of white dwarfs with magnetic fields} 

The core of a white dwarf consists of a dense Coulomb plasma of nuclei in a charge compensating  background of relativistic electrons. Apart from carbon and oxygen (the primary ashes of helium burning), the core may contain other nuclei like helium~\cite{nelemans1998,liebert2004,benvenuto2005}, neon and magnesium~\cite{nomoto1984}, or even iron~\cite{panei2000,catalan2008}. Iron white dwarfs could be formed from the explosive ignition of electron degenerate oxygen-neon-magnesium cores~\cite{isern1991}, or from failed detonation supernovae~\cite{jordan2012}. For simplicity, we assume that the stellar core is made of only one type of nuclei with charge number $Z$ and mass number $A$ (see, e.g. Ref.~\citenum{fantina2015} for the treatment of mixtures). We further suppose that the star has  sufficiently cooled down such that thermal effects can be neglected. 

Whereas nuclei with number density $n_N$ exert a negligible pressure $P_N=0$, they contribute to the mass density
\begin{equation}
	\label{eq:rho}
	\rho=n_N M^\prime(A,Z)\, ,
\end{equation}
where $M^\prime(A,Z)$ denotes the ion mass including the rest mass of $Z$ electrons and can be obtained from the experimental \emph{atomic} mass $M(A,Z)$ by subtracting out the binding energy of the atomic electrons (see Eq.~(A4) of Ref.~\citenum{lunney2003}). In principle, the presence of a magnetic field can have some effect on nuclei~\cite{arteaga2011,stein2016}. However, the change of nuclear masses is negligible even for the strongest magnetic fields of order $10^{15}$~G expected in super Chandrasekhar white dwarfs~\cite{subramanian2015,bera2016,chatterjee2017}.

To a very good approximation, electrons can be treated as an ideal Fermi gas. In the presence of a  magnetic field, the electron motion perpendicular to the field is quantized into Landau-Rabi  
levels~\cite{rabi1928,landau1930}. Quantization effects on the equation of state are significant when $B$ exceeds $B_\text{rel}$, where 
\begin{equation}\label{eq:Bcrit}
	B_\text{rel}\equiv \frac{m_e^2 c^3}{e\hbar}\approx 4.4\times 10^{13}~\text{G}\, ,
\end{equation} 
where $m_e$ is the electron mass and $e$ is the elementary electric charge. 
The expressions for the energy density $\mathcal{E}_e$ and pressure $P_e$ for arbitrary magnetic 
field strength can be found in Ref.~\citenum{haensel2007}. 

The main correction to the ideal electron Fermi gas arises from the electron-ion interactions, as first shown by Frenkel~\cite{frenkel1928} (see e.g. Ref.~\citenum{haensel2007} for a discussion of higher-order
corrections).   
For pointlike ions embedded in a uniform electron gas with number density $n_e=Z n_N$ (from electric charge neutrality), 
the corresponding energy density is given by (see e.g. Chap. 2 of Ref.~\citenum{haensel2007})
\begin{equation}\label{eq:EL}
	\mathcal{E}_L=C_M  \left(\frac{4\pi}{3}\right)^{1/3} e^2 n_e^{4/3} Z^{2/3}\, ,
\end{equation}
where $C_M$ is the Madelung constant. 
The contribution to the pressure is thus given by  
\begin{equation}\label{eq:PL}
	P_L=n_e^2 \frac{d(\mathcal{E}_L/n_e)}{dn_e}=\frac{\mathcal{E}_L}{3}\, . 
\end{equation}
The pressure of the Coulomb plasma finally reads $P=P_e+P_L$.

In the following, we shall consider that ions are arranged in a body-centered cubic lattice since this configuration leads to the lowest energy~\cite{haensel2007}. 
In this case, the Madelung constant is given by~\cite{baiko2001} $C_M=-0.895929255682$. 
According to the Bohr-van Leeuwen theorem~\cite{vanvleck1932}, the electrostatic corrections \eqref{eq:EL} and \eqref{eq:PL} are independent 
of the magnetic field apart from a negligibly small contribution due to quantum zero-point motion of ions about 
their equilibrium position~\cite{baiko2009}.

\section{Onset of electron captures by nuclei in dense environments}

The onset of electron captures by nuclei $(A,Z)$ is formally determined by the same condition 
irrespective of the magnetic field strength by requiring the constancy of the Gibbs free energy per nucleon at fixed temperature and pressure~\cite{fantina2015}. The threshold electron Fermi energy is found to first order in the fine-structure constant $\alpha=e^2/(\hbar c)$ from the condition:
\begin{equation}\label{eq:e-capture-gibbs-approx}
	\gamma_e + C_M \left(\frac{4\pi n_e}{3}\right)^{1/3} \alpha \lambda_e F(Z) = \gamma_e^{\beta}(A,Z) \, ,
\end{equation}
\begin{equation}\label{eq:def-F}
	F(Z)\equiv Z^{5/3}-(Z-1)^{5/3} + \frac{1}{3} Z^{2/3}\, ,
\end{equation}
\begin{equation}\label{eq:muebeta}
	\gamma_e^{\beta}(A,Z)\equiv -\frac{Q_{\rm EC}(A,Z)}{m_e c^2} + 1 \, ,
\end{equation}
where $\gamma_e\equiv \mu_e/(m_e c^2)$, $\lambda_e=\hbar /(m_e c)$ is the electron Compton wavelength,
and we have introduced the $Q$-value (in vacuum) associated with electron capture by nuclei ($A,Z$): 
\begin{equation}
	Q_{\rm EC}(A,Z) = M^\prime(A,Z)c^2-M^\prime(A,Z-1)c^2\, .
\end{equation}
These $Q$-values can be obtained from the tabulated $Q$-values of $\beta$ decay by the following relation:
\begin{equation}
	Q_{\rm EC}(A,Z) = -Q_\beta(A,Z-1)\, .
\end{equation}
In principle, the daughter nucleus may be in an excited state. However, such a transition would occur at a higher density. 

In the absence of magnetic fields, the threshold condition~(\ref{eq:e-capture-gibbs-approx}) can be solved analytically~\cite{chamelfantina2016}. Recalling that the electron Fermi energy  is given by 
\begin{equation}
	\mu_e = m_e c^2 \sqrt{1+x_r^2}\, , 
\end{equation}
where  $x_r=\lambda_e k_e$ and $k_e=(3\pi^2 n_e)^{1/3}$ is the electron Fermi wave number, 
the solution reads 
\begin{eqnarray}\label{eq:exact-xr}
	x^\beta_r=\gamma_e^{\beta} \Biggl\{\sqrt{1-\biggl[1-\tilde F(Z)^2\biggr]/(\gamma_e^{\beta})^{2}}
	-\tilde F(Z)\Biggr\} \Biggl[1-\tilde F (Z)^2\Biggr]^{-1}\, ,
\end{eqnarray}
with 
\begin{equation}\label{eq:Ftilde}
	\tilde F(Z)\equiv C_M \left(\frac{4}{9\pi}\right)^{1/3}\alpha F(Z)\, .
\end{equation}
The pressure $P_{\beta}(A,Z)$ at the onset of electron captures is 
given by
\begin{eqnarray}
	\label{eq:exact-Pbeta}
	P_{\beta}(A,Z)&=&\frac{m_e c^2}{8 \pi^2 \lambda_e^3} \biggl[x^\beta_r\left(\frac{2}{3}(x^\beta_r)^2-1\right)\sqrt{1+(x^\beta_r)^2}+\ln(x^\beta_r+\sqrt{1+(x^\beta_r)^2})\biggr]\nonumber\\
	&&+\frac{C_M \alpha}{3}\left(\frac{4}{243\pi^7}\right)^{1/3} (x^\beta_r)^4 \frac{m_e c^2}{\lambda_e^3}Z^{2/3} \, .
\end{eqnarray}
The corresponding average mass density is found from Eq.~\eqref{eq:rho} and is given by 
\begin{eqnarray}\label{eq:exact-rhobeta}
	\rho_\beta(A,Z)  = \frac{M^\prime(A,Z)}{Z}  \frac{(x^\beta_r)^3 }{3 \pi^2 \lambda_e^3}\, .
\end{eqnarray}
Numerical results are summarized in Table~\ref{tab1} for nuclei expected to be found in the core of white dwarfs, using data from the 2020 Atomic Mass Evaluation~\cite{wang2021}. Fundamental constants were taken from NIST CODATA 2018\footnote{\url{https://physics.nist.gov/cuu/Constants/index.html}}. We have not considered  helium since this element is expected to undergo pycnonuclear fusion in white-dwarf cores before capturing electrons.

\begin{table}
	\tbl{Dimensionless threshold electron Fermi energy $\gamma_e=\mu_e/(m_e c^2)$, mass density $\rho_\beta$ and pressure $P_\beta$ for the onset of electron captures by the given nuclei in unmagnetized white dwarfs.}
	{\begin{tabular}{@{}cccccc@{}}
			\toprule
			             & $^{16}$O              & $^{12}$C              & $^{20}$Ne            & $^{24}$Mg             & $^{56}$Fe \\
			\colrule
			$\gamma_e$                 & $22.0$               & $27.8$               & $15.2$              & $12.2$               & $8.74$     \\
			$\rho_\beta$ [g~cm$^{-3}$] & $2.06\times 10^{10}$ & $4.16\times 10^{10}$ & $6.81\times 10^9$   & $3.51\times 10^9$    & $1.37 \times10^9$ \\
			$P_\beta$ [dyn~cm$^{-2}$]     & $2.73\times 10^{28}$ & $6.99\times 10^{28}$ & $6.21\times10^{27}$ & $2.56\times 10^{27}$ & $6.46\times10^{26}$     \\
			\botrule
		\end{tabular}
	}
	\label{tab1}
\end{table}

In the presence of a magnetic field, the threshold condition~(\ref{eq:e-capture-gibbs-approx})
must be solved using the following relation between $n_e$ and $\gamma_e$:
\begin{equation}\label{eq:ne}
	n_e =\frac{2 B_\star}{(2 \pi)^2 \lambda_e^3} \sum_{\nu=0}^{\nu_{\rm  max}} g_\nu x_e(\nu)\, ,
\end{equation}
\begin{equation}\label{eq:xe}
	x_e(\nu) =\sqrt{\gamma_e^2 -1-2 \nu B_\star}\, ,
\end{equation}
where we have introduced $B_\star\equiv B/B_\text{rel}$, 
and the degeneracy $g_\nu=1$ for $\nu=0$ and $g_\nu=2$ for $\nu\geq 1$. The index $\nu_\text{max}$ is the highest integer for which $\gamma_e^2-1-2\nu_{\rm  max}B_\star\geq 0$, i.e. 
\begin{equation}
\nu_\text{max}=\left[\frac{\gamma_e^2-1}{2 B_\star}\right]\, , 
\end{equation}
where $[.]$ denotes the integer part. 

In the weakly quantizing regime meaning that many Landau-Rabi levels are populated ($\nu_\text{max}\gg 1$), analytical solutions can be found. Remarking that the magnetic field enters explicitly in  \eqref{eq:e-capture-gibbs-approx} only through the small electrostatic correction, the threshold electron Fermi energy is still approximately given by the solution in the absence of magnetic fields, namely $\gamma_e\approx \sqrt{1+(x^\beta_r)^2}$ with $x^\beta_r$ given by \eqref{eq:exact-xr}. Substituting in Eqs.~\eqref{eq:ne} and \eqref{eq:xe}, using the expansions (41) obtained in Ref.~\citenum{dib2001}
leads to the following estimate for the density marking the onset of electron captures: 
\begin{equation}\label{eq:rhobeta-mag-weak}
\rho_\beta(A,Z)\approx \frac{M^\prime(A,Z)}{2\pi^2 Z \lambda_e^3} \bigg[ \frac{2}{3}\left(\gamma_e^2-1\right)^{3/2}+(2B_\star)^{3/2}\zeta\left(\frac{-1}{2},\left\{\frac{\gamma_e^2-1}{2B_\star}\right\}\right)+\frac{B_\star^2}{6\sqrt{\gamma_e^2-1}}\biggr]\, , 
\end{equation} 
where $\zeta(z,q)$ is the Hurwitz zeta function defined by 
\begin{equation}
	\zeta(z,q)=\sum_{\nu=0}^{+\infty} \frac{1}{(\nu+q)^z}
\end{equation} 
for $\Re(z)>1$ and by analytic continuation to other $z\neq 1$ (excluding poles $\nu+q=0$), 
and $\{.\}$ in the argument denotes the fractional part. The first term in Eq.~\eqref{eq:rhobeta-mag-weak} represents the threshold density in the absence of magnetic field. The second term accounts for magnetic oscillations while the last term is a higher-order correction. The expression for the associated pressure is more involved. Using Eqs.~(41), (43) and (44) of Ref.~\citenum{dib2001} yields\footnote{In the notations of Ref.~\citenum{dib2001}, the electron contribution to the pressure can be directly obtained from the grand potential density by $P_e=-\omega_0^{(\rm mon)}-\omega_0^{(\rm osc)}$. The total pressure is found by adding the electrostatic correction~\eqref{eq:PL} using the expansion for the electron density.}  
\begin{align}\label{eq:Pbeta-mag-weak}
P_\beta(A,Z)& \approx \frac{m_e c^2}{4\pi^2 \lambda_e^3}\biggl\{\frac{1}{2}\left(1-2B_\star+\frac{2B_\star^2}{3}\right)\log\left(\frac{\gamma_e+\sqrt{2B_\star+\gamma_e^2-1}}{1+\sqrt{2B_\star}}\right)\nonumber \\ &-\frac{1}{2}\left(\gamma_e\sqrt{2B_\star+\gamma_e^2-1}-\sqrt{2B_\star}\right)+\frac{1}{3}\left(\gamma_e\sqrt{2B_\star+\gamma_e^2-1}^3-\sqrt{2B_\star}^3\right)\nonumber \\ 
&+B_\star \left({\rm arccosh}~\gamma_e - \gamma_e\sqrt{\gamma_e^2-1}\right)-(2B_\star)^{5/2} \int_0^{+\infty}\frac{\tilde{\zeta}_3(-1/2,q+1)}{\sqrt{1+2B_\star q}} dq \nonumber \\
&+ \frac{2}{3} \frac{(2 B_\star)^{5/2}}{\gamma_e}\zeta\left(\frac{-3}{2},\left\{\frac{\gamma_e^2-1}{2B_\star}\right\}\right)  
+\frac{2}{15}\frac{(2B_\star)^{7/2}}{\gamma_e^3}\zeta\left(\frac{-5}{2},\left\{\frac{\gamma_e^2-1}{2B_\star}\right\}\right)  \nonumber \\
&+\frac{1}{240}\left(\frac{B_\star}{\gamma_e}\right)^4 +4B_\star^2 \int_0^1 \zeta\left(\frac{-1}{2},q\right)\zeta\left(\frac{1}{2},q+\frac{1}{2B_\star}\right)dq \nonumber \\ 
	&+\frac{2}{3}\left(\frac{2}{3\pi}\right)^{1/3} C_M \alpha Z^{2/3}\biggl[  \frac{2}{3}\left(\gamma_e^2-1\right)^{3/2} \nonumber \\ 
	&+(2B_\star)^{3/2}\zeta\left(\frac{-1}{2},\left\{\frac{\gamma_e^2-1}{2B_\star}\right\}\right)+\frac{B_\star^2}{6\sqrt{\gamma_e^2-1}}\biggr]^{4/3} \biggr\}
	\, ,
\end{align}
with 
\begin{equation}
	\tilde{\zeta}_3(z,q)=\zeta(z,q)-\frac{1}{z-1}q^{-z+1}-\frac{1}{2}q^{-z}-\frac{z}{12}q^{-z-1}\, .
\end{equation}
An analytical solution also exists in the strongly quantizing regime whereby electrons are all confined to the lowest Landau-Rabi level ($\nu_\text{max}=0$). Introducing 
\begin{equation}
	\bar F(Z,B_\star)\equiv \frac{1}{3} C_M \alpha F(Z)\left(\frac{2 B_\star}{3\pi}\right)^{1/3} <0\, ,
\end{equation}
\begin{equation}
	\upsilon\equiv \frac{\gamma_e^{\beta}}{2 |\bar F(Z, B_\star)|^{3/2}}\, ,
\end{equation}
the solutions are given by the following formulas~\cite{chamel2020}: 
\begin{equation}\label{eq:exact-gammae}
	\gamma_e=\begin{cases}
		8|\bar F(Z, B_\star)|^{3/2}\, {\rm cosh}^3\left(\frac{1}{3}{\rm arccosh\,} \upsilon\right) & \text{if} \ \upsilon\geq 1\, ,\\
		8|\bar F(Z, B_\star)|^{3/2}\, \cos^3\left( \frac{1}{3}\arccos \upsilon\right) & \text{if} \ 0\leq \upsilon< 1\, . 
	\end{cases}
\end{equation}
The threshold pressure and density are respectively given by: 
\begin{align}\label{eq:exact-Pbeta-mag}
	P_{\beta}(A,Z,B_\star) &=\frac{B_\star m_e c^2 }{4 \pi^2 \lambda_e^3 }\biggl[\gamma_e\sqrt{\gamma_e^2-1}-\ln\left(\sqrt{\gamma_e^2-1}+\gamma_e\right) \nonumber \\ 
	&+\frac{C_M \alpha }{3}\left(\frac{16 B_\star Z^2 }{3\pi}\right)^{1/3} \left(\gamma_e^2-1\right)^{2/3} \biggr] \, ,
\end{align} 
\begin{equation}\label{eq:exact-rhobeta-mag}
	\rho_\beta(A,Z,B_\star) = \frac{B_\star}{2\pi^2 \lambda_e^3} \frac{M^\prime(A,Z)}{Z}   \sqrt{\gamma_e^2-1} \, . 
\end{equation}
Let us recall that Eq.~\eqref{eq:exact-gammae} is only valid if all electrons lie in the lowest 
Landau-Rabi level, i.e. if their Fermi energy does not exceed $\gamma_e=\sqrt{1+2B_\star}\approx \sqrt{2B_\star}$ or equivalently if their density does not exceed $n_e=B_{\star}^{3/2}/(\sqrt{2}\pi^2\lambda_e^3)$ using Eqs.~\eqref{eq:ne} and \eqref{eq:xe}. This condition translates into a lower bound for the magnetic field $B_\star \geq B_{\star1}$. 
To find $B_{\star1}$, we substitute the above expressions for $\gamma_e$ and $n_e$ in Eq.~\eqref{eq:e-capture-gibbs-approx}. This leads to 
\begin{equation}\label{eq:Bstar1}
B_{\star1}=\frac{(\gamma_e^\beta)^2}{2}\biggl[1+\frac{C_M\alpha}{(3\pi)^{1/3}} F(Z)
\biggr]^{-2}\, .
\end{equation}
Values for some nuclei are summarized in Table~\ref{tab2}. For magnetic field strength $B_\star\lesssim B_{\star1}$, the threshold condition~(\ref{eq:e-capture-gibbs-approx}) must be solved numerically. 

\begin{table}
	\tbl{Magnetic field strength in units of $B_\text{rel}$ above which electrons are confined to the lowest Rabi level when captured by the given nuclei.}
	{\begin{tabular}{@{}ccccc@{}}
			\toprule
			$^{16}$O & $^{12}$C & $^{20}$Ne & $^{24}$Mg & $^{56}$Fe \\
			\colrule
			240 & 384 & 115 & 74.1 & 37.8 \\
			\botrule
		\end{tabular}
	}
	\label{tab2}
\end{table}

\begin{figure}[h]%
	\begin{center}
		\parbox{2.3in}{\includegraphics[width=2.4in]{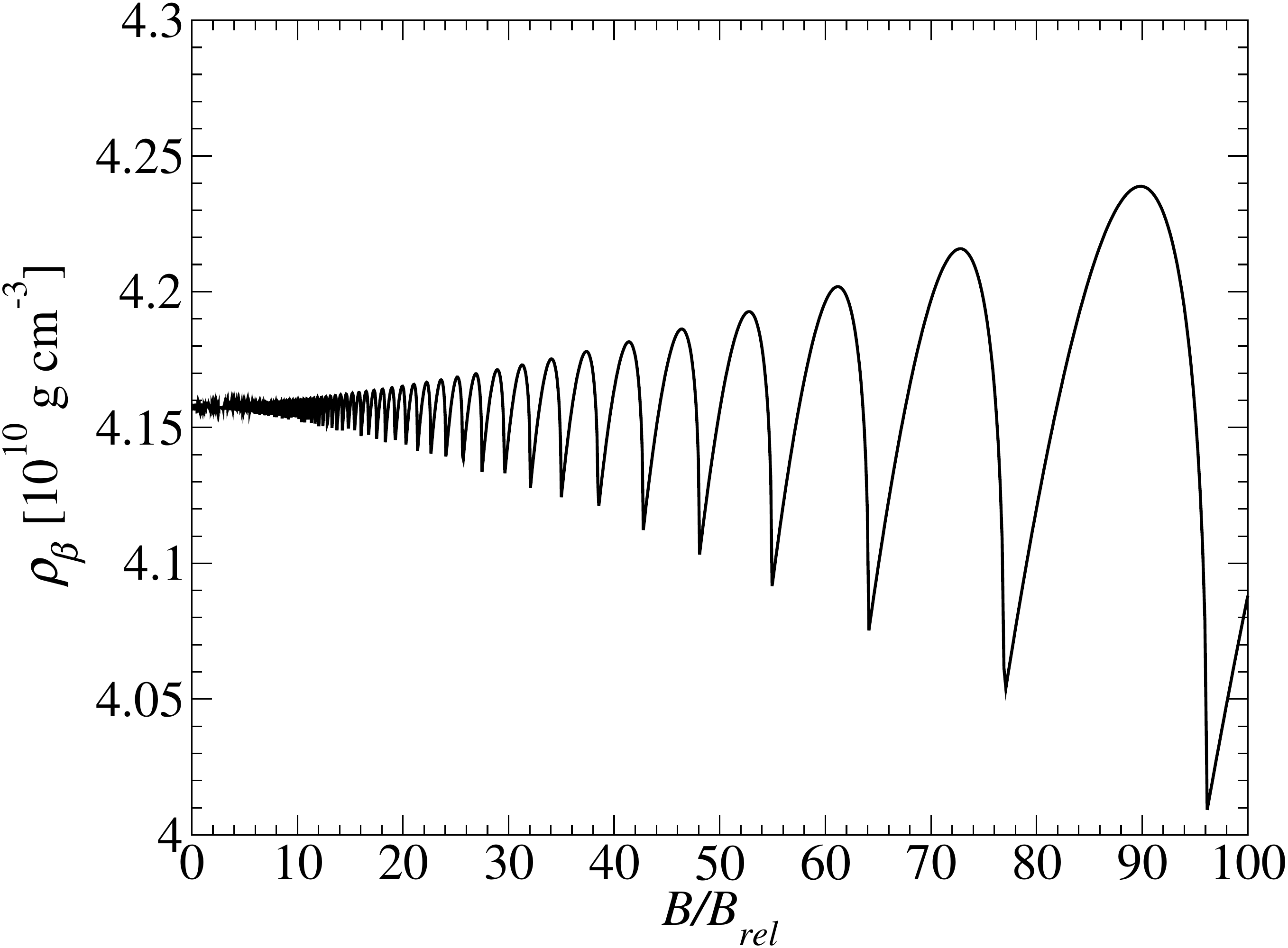}\figsubcap{a}}
		\hspace*{4pt}
		\parbox{2.3in}{\includegraphics[width=2.4in]{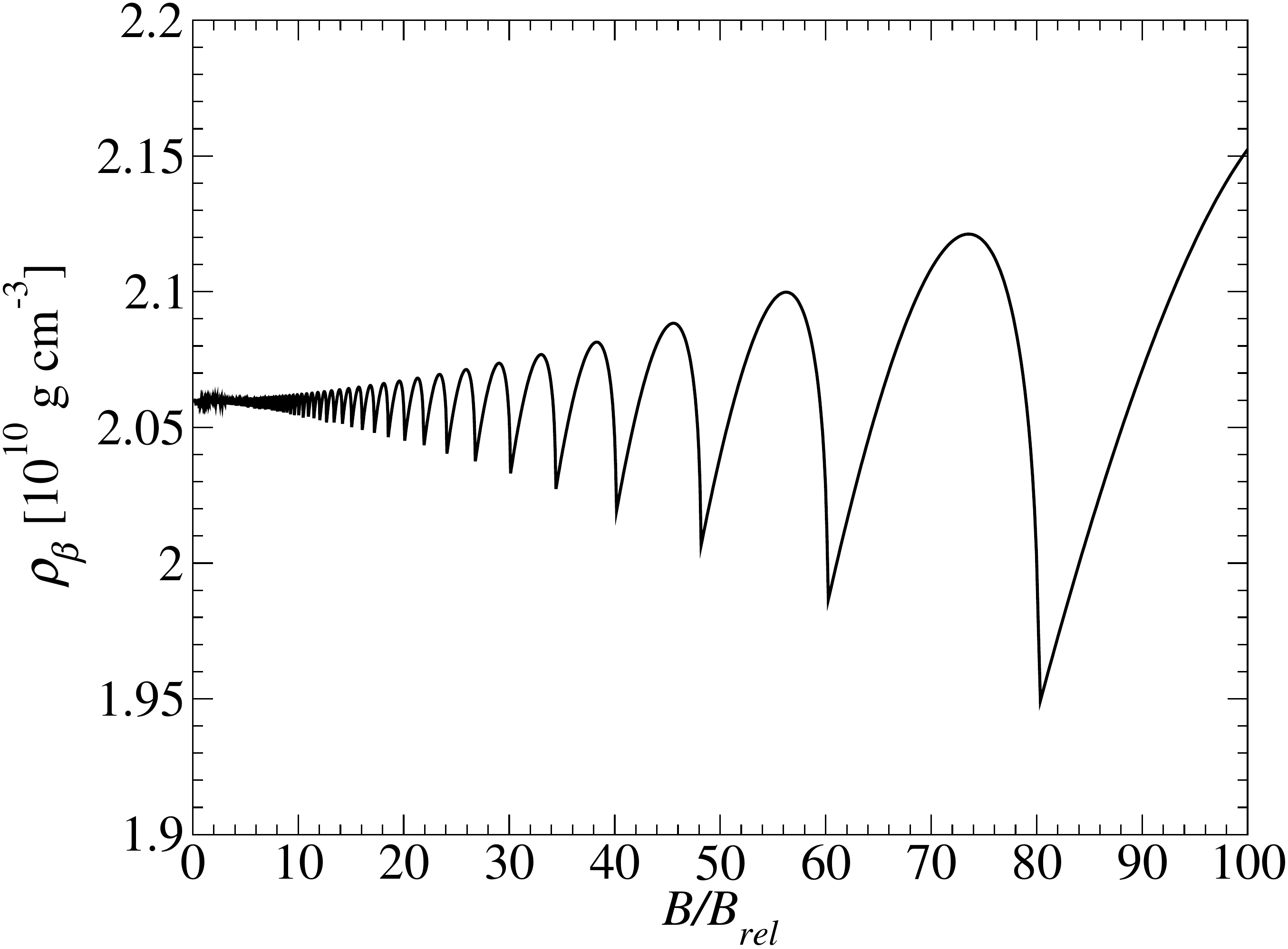}\figsubcap{b}}
		\caption{Threshold density for the onset of electron captures by nuclei as a function of the magnetic field strength in units of the characteristic field~\eqref{eq:Bcrit}. 
		(a) Results for $^{12}$C. (b) Results for $^{16}$O.}%
		\label{fig1}
	\end{center}
\end{figure}

\begin{figure}[h]%
	\begin{center}
		\parbox{2.3in}{\includegraphics[width=2.4in]{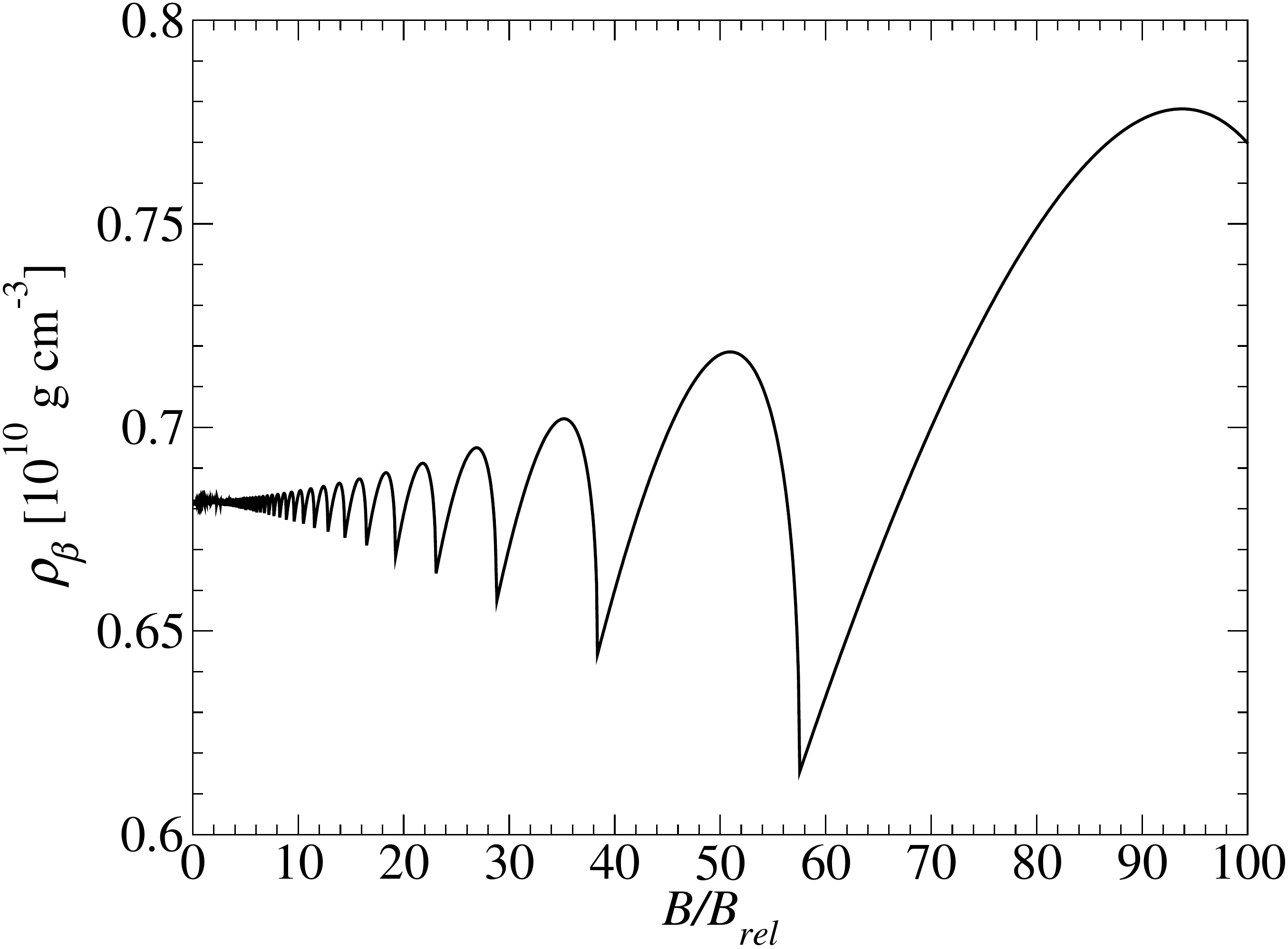}\figsubcap{a}}
		\hspace*{4pt}
		\parbox{2.3in}{\includegraphics[width=2.4in]{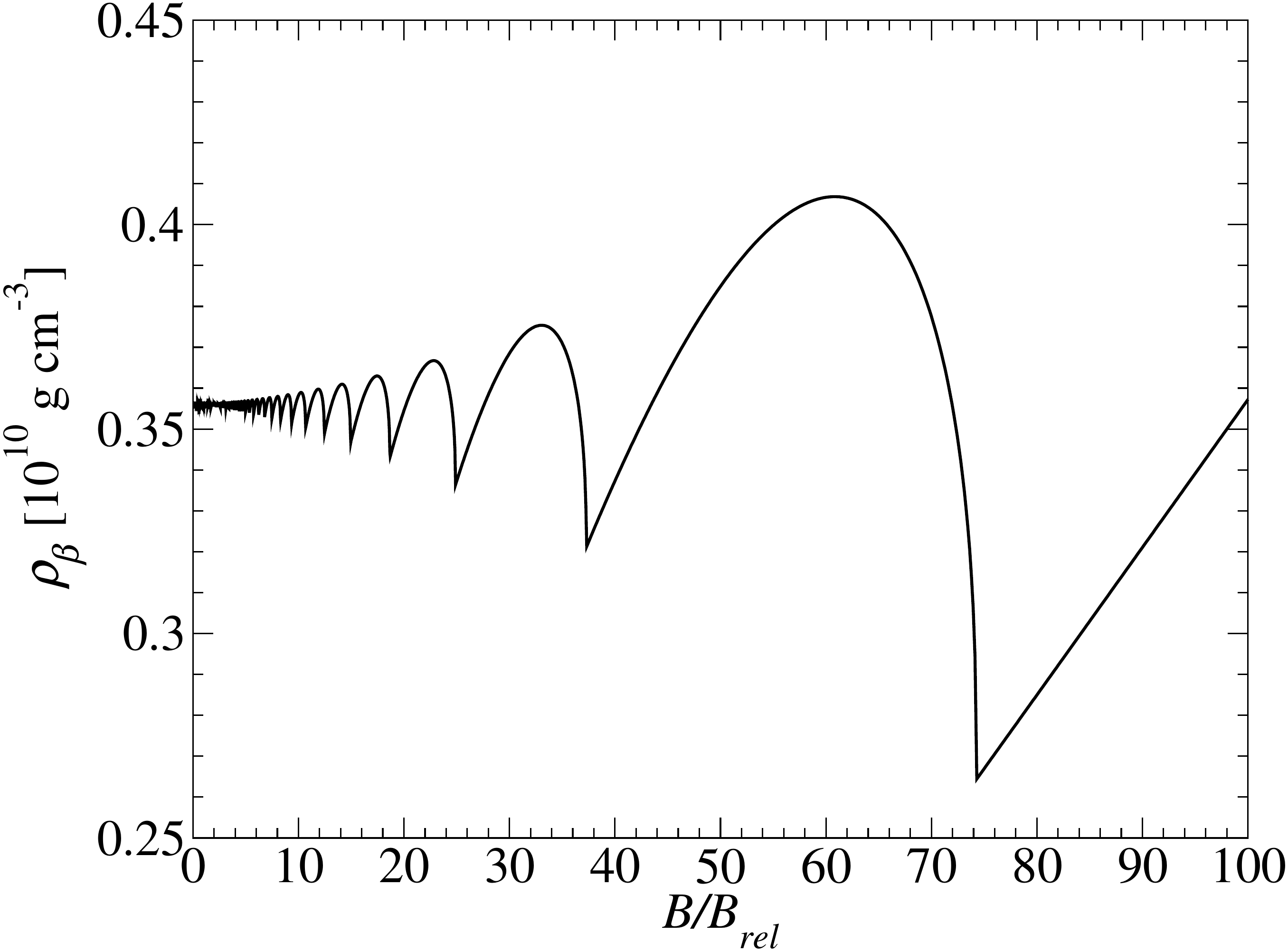}\figsubcap{b}}
		\caption{Same as Fig.~\ref{fig1} for (a) $^{20}$Ne and (b) $^{24}$Mg.}%
		\label{fig2}
	\end{center}
\end{figure}

Full results for different nuclei expected to be found in white-dwarf cores are shown in Figs.~\ref{fig1} and \ref{fig2}. Changes in the occupation of Rabi levels as the magnetic field is increased lead to typical oscillations of the threshold density $\rho_\beta$: the onset of electron captures can thus be shifted to either lower or higher density as compared to unmagnetized matter. The lowest value is reached for $B_\star=B_{\star1}$ and lies about $25\%$ below the threshold density in the absence of magnetic fields irrespective of the composition (this can  be easily shown by taking the  ratio of Eqs.~\eqref{eq:exact-rhobeta-mag} and \eqref{eq:exact-rhobeta} using \eqref{eq:Bstar1} and ignoring all terms in $\alpha$). In the strongly quantizing regime, $\rho_\beta$ is unbound from above and increases almost linearly with $B_\star$. 

\section{Global structure of white dwarfs}

The core of a white dwarf is expected to be surrounded by a helium mantle and an hydrogen envelope. 
Their contribution to the mass of the star cannot exceed $\sim 1\%$ to avoid a thermonuclear runaway. For simplicity, we will ignore these layers here. 

In the model originally considered by Chandrasekhar~\cite{chandra1931}, the interior of a white dwarf was described using the equation of state of an ideal electron Fermi gas. The composition, assumed to be uniform, was only included throughout the charge neutrality condition. In the ultrarelativistic limit, the resulting equation of state reduces to the polytropic form  $P\approx K_0 \rho^{4/3}$ with 
\begin{equation}\label{eq:K0}
	K_0=\frac{\hbar c (3\pi^2)^{1/3}}{4} \left(\frac{Z}{M^\prime(A,Z)}\right)^{4/3}\, .
\end{equation}
The mass and radius of a white dwarf with central density $\rho_c$ can be deduced from the theory of polytropes~\cite{chandra1957}
\begin{equation}\label{eq:Mpolytrope}
	M_\text{Ch} =4\pi \left(\frac{K_0}{\pi G}\right)^{3/2} \xi_1^2 \vert\theta^\prime(\xi_1)\vert\, ,
\end{equation}
\begin{eqnarray}\label{eq:Rpolytrope}
	R_\text{Ch}=\sqrt{\frac{K_0}{\pi G}}\frac{\xi_1}{\rho_c^{1/3}} 
\end{eqnarray}
respectively, where $\xi_1\simeq 6.89685$ and $\xi_1^2 \vert\theta^\prime(\xi_1)\vert\simeq 2.01824$. 
Substituting the expression for $K_0$ thus yields
\begin{equation}\label{eq:Mch}
M_\textrm{Ch} =\frac{\sqrt{3\pi}}{2} \xi_1^2 \vert\theta^\prime(\xi_1)\vert \frac{m_\text{P}^3}{m^2}  y_e^2 \simeq  3.09798 \frac{m_\text{P}^3}{m^2}  y_e^2\, ,
\end{equation}
\begin{eqnarray}\label{eq:Rch}
R_\text{Ch}=\frac{\sqrt{3\pi}}{2}\xi_1 \lambda_e \frac{m_\text{P}}{m} \frac{y_e}{x_{r,c}}\simeq
10.5866 \lambda_e \frac{m_\text{P}}{m} \frac{y_e}{x_{r,c}}  \, ,
\end{eqnarray}
where 
\begin{equation}
m\equiv \frac{M^\prime(A,Z)}{A}\, ,
\end{equation}
and $x_{r,c}$ denotes the relativity parameter at the density $\rho_c$. 
Equation~\eqref{eq:Mch} coincides with the maximum mass because the ultrarelativistic regime 
is only valid at high densities such that the relativity parameter in the central core of the star satisfies $x_{r,c}\gg 1$. 
In the limit $x_{r,c}\rightarrow +\infty$, the radius $R_\text{Ch}$ vanishes. 

The electrostatic corrections \eqref{eq:EL} and \eqref{eq:PL} can be easily taken into account by simply renormalizing the constant $K_0$ as 
\begin{equation}\label{eq:K}
	K = K_0 \biggl[1+ \alpha \frac{8 C_M}{6}\left(\frac{4}{9\pi}\right)^{1/3} Z^{2/3} \biggr]\, .
\end{equation}
The mass and radius thus become\footnote{Such kind of scaling was briefly mentioned in Ref.~\citenum{hamada1961} through an effective mean molecular weight per electron.}
\begin{equation}\label{eq:M}
	M= \biggl[1+ \alpha \frac{8 C_M}{6}\left(\frac{4}{9\pi}\right)^{1/3} Z^{2/3} \biggr]^{3/2} M_\text{Ch} < M_\textrm{Ch} \, ,
\end{equation}
\begin{eqnarray}\label{eq:R}
	R=R_\text{Ch}  \biggl[1+ \alpha \frac{8 C_M}{6}\left(\frac{4}{9\pi}\right)^{1/3} Z^{2/3} \biggr]^{1/2} < R_\textrm{Ch} \, .
\end{eqnarray}
This shows how the electron-ion interactions lower both the mass and the radius. 
It can be seen that the radius of the most massive white dwarfs still vanishes in the limit $x_{r,c}\rightarrow +\infty$. However, as discussed earlier, nuclei in the white dwarf core 
will capture electrons as soon as the central density  $\rho_c$  exceeds the threshold density
$\rho_\beta$. The daughter nuclei are generally unstable against a second electron capture. 
The transition is accompanied by a discontinuous increase of density given by
\begin{equation}
	\frac{\Delta  \rho}{ \rho_\beta} = \frac{Z}{Z-2}\frac{M^\prime(A,Z-2)}{M^\prime(A,Z)}\biggl[1+\alpha C_M\left(\frac{4}{9\pi}\right) ^{1/3}\left(Z^{2/3}-(Z-2)^{2/3}\right)\frac{\sqrt{1+x_r^2}}{x_r}\biggr]-1\, ,
\end{equation}
where $x_r$ is given by Eq.~\eqref{eq:exact-xr}. The density jump is about 50\% for carbon, 33\% for oxygen, 25\% for neon, 20\% for magnesium, and 8.2\% for iron. 

Because these reactions occur at the same 
pressure, the adiabatic index defined by 
\begin{equation}
\Gamma=\frac{d\log P}{d\log{\rho}}
\end{equation}
therefore vanishes thus making the star unstable. In the limit of ultrarelativistic electrons, the average threshold density and pressure reduce to
\begin{equation}\label{eq:rhobeta-ultra}
	\rho_\beta(A,Z) \approx \frac{M^\prime(A,Z)}{Z}  \frac{\mu_e^\beta(A,Z)^3}{3\pi^2 (\hbar c)^3}
	\biggl[1+\alpha C_M  \left(\frac{4}{9\pi}\right)^{1/3}F(Z)\biggr]^{-3}\, ,
\end{equation}
\begin{equation}\label{eq:Pbeta}
	P_\beta(A,Z) \approx \frac{\mu_e^\beta(A,Z)^4}{12 \pi^2 (\hbar c)^3}\biggl[1+
	4 C_M \alpha Z^{2/3} \left(\frac{4}{243 \pi}\right)^{1/3}
\biggr] \biggl[1+\alpha C_M  \left(\frac{4}{9\pi}\right)^{1/3}F(Z)\biggr]^{-4}\, .
\end{equation}
Substituting Eq.~\eqref{eq:rhobeta-ultra} in \eqref{eq:R} using \eqref{eq:Rch} leads to the following lower bound for the radius of a white dwarf: 
\begin{eqnarray}\label{eq:Rmin}
	R_\text{min}=\frac{\sqrt{3\pi}}{2}\xi_1 \lambda_e \frac{m_\text{P}}{m} \frac{y_e}{\gamma_e^\beta(A,Z)}\biggl[1+\alpha C_M\left(\frac{4}{9\pi}\right)^{1/3}F(Z)\biggr]\nonumber \\ 
	\times \biggl[1+ \alpha \frac{8 C_M}{6}\left(\frac{4}{9\pi}\right)^{1/3} Z^{2/3} \biggr]^{1/2} \, .
\end{eqnarray}
The factors in square brackets account for the electron-ion interactions:  the first arises 
from the shift in the threshold density for the onset of electron captures whereas the second is due to the correction to the 
equation of state. Both factors lead to a reduction of the radius. Results of Eq.~\eqref{eq:Rmin} are summarized in  Table~\ref{tab3}. 

\begin{table}
	\tbl{Minimum radius (in km) of nonmagnetic stable white dwarfs.}
	{\begin{tabular}{@{}cccccc@{}}
			\toprule
			$^{12}$C    & $^{16}$O   & $^{20}$Ne  & $^{24}$Mg & $^{56}$Fe   \\
			\colrule
            $958$        & $1209$     & $1743$      & $2168$      &  $2791$ \\
			\botrule
		\end{tabular}
	}
	\label{tab3}
\end{table}

The mass remains unchanged since it does not depend on the density and is still given by Eq.~\eqref{eq:M}. However, departure from the polytropic equation of state $P=K\rho^{4/3}$ and the fact that the central density $\rho_c$ is finite induce a slight reduction of the maximum mass. Following the perturbative approach described in 
Chapter 6 of Ref.~\citenum{shapiro1983} and based on the minimization of some approximation for the total energy of the star, we find 
\begin{eqnarray}
	\frac{\delta M}{M}\approx -\frac{k_3}{k_2}\frac{3\pi}{(\gamma_e^\beta)^2}\left(\frac{3}{2\xi_1^4 \vert\theta^\prime(\xi_1)\vert^2 }\right)^{1/3}\biggl[1+ \alpha \frac{8 C_M}{6}\left(\frac{4}{9\pi}\right)^{1/3}Z^{2/3}\biggr]^{-1}	\nonumber \\ 
	\times \biggl[1+\alpha C_M  \left(\frac{4}{9\pi}\right)^{1/3}F(Z)\biggr]^{2}
\end{eqnarray}
where $k_2=0.639001$ and $k_3=0.519723$. Results are summarized in Table~\ref{tab4}. The influence of electron captures and electron-ion interactions are all the more important that matter contains heavier elements: whereas the overall reduction of the Chandrasekhar mass $M_\text{Ch}$ amounts to 3\% for carbon, it reaches 13\% for iron. 

\begin{table}
	\tbl{Maximum mass (in solar units) of nonmagnetic white dwarfs for an ultrarelativistic electron Fermi gas (first line), with electrostatic correction (second line) and electron captures (third line).}
	{\begin{tabular}{@{}cccccc@{}}
			\toprule
			$^{12}$C    & $^{16}$O & $^{20}$Ne & $^{24}$Mg & $^{56}$Fe   \\
			\colrule
			$1.456$     & $1.457$  & $1.457$   & $1.458$   & $1.259$ \\
			$1.424$     & $1.418$  & $1.411$   & $1.406$   & $1.184$ \\ 
			$1.413$     & $1.401$  & $1.377$   & $1.353$   & $1.095$ \\
			\botrule
		\end{tabular}
	}
	\label{tab4}
\end{table}

\begin{figure}[h]%
	\begin{center}
		\parbox{2.4in}{\includegraphics[width=2.6in]{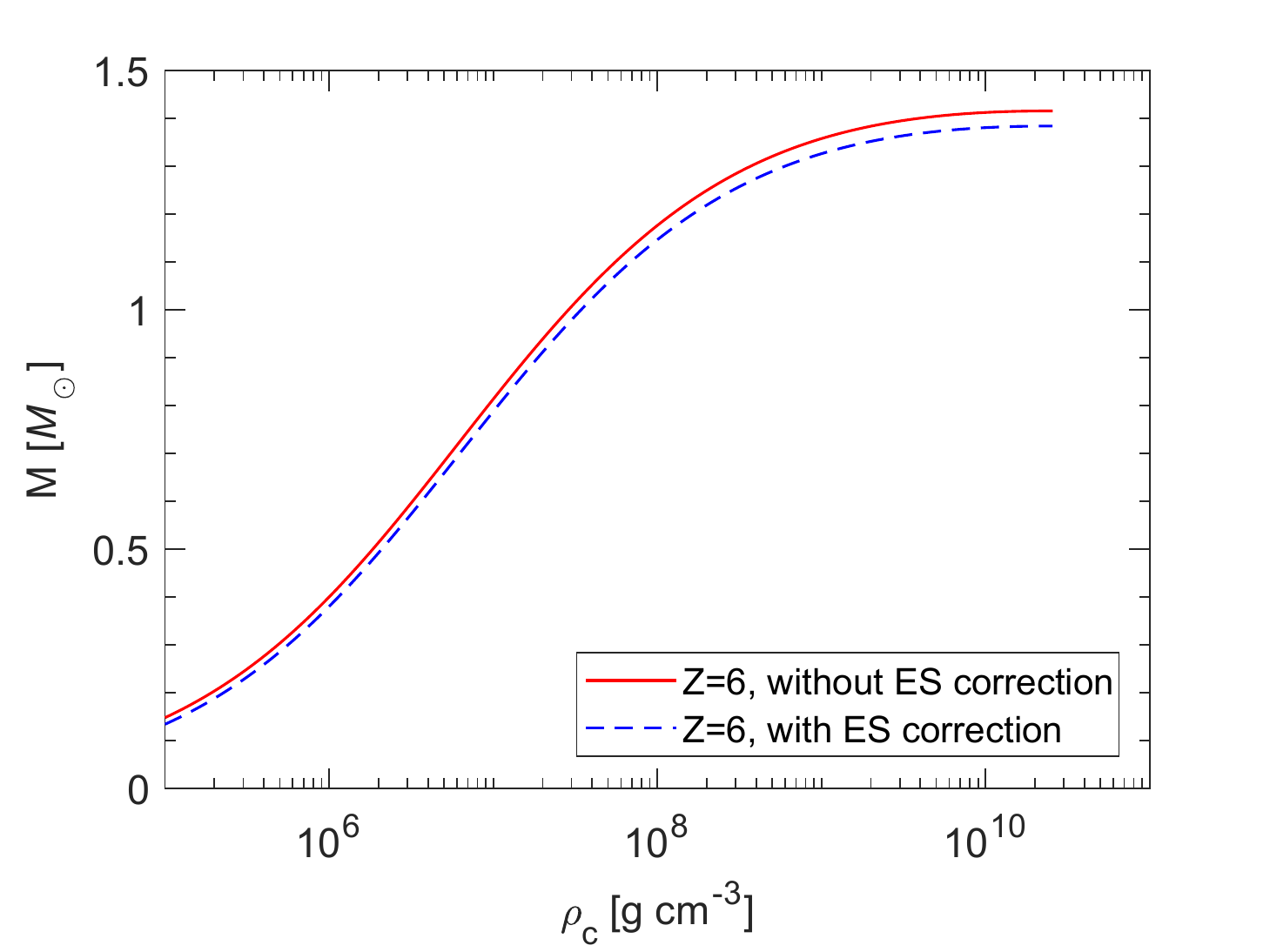}\figsubcap{a}}
		\hspace*{4pt}
		\parbox{2.4in}{\includegraphics[width=2.6in]{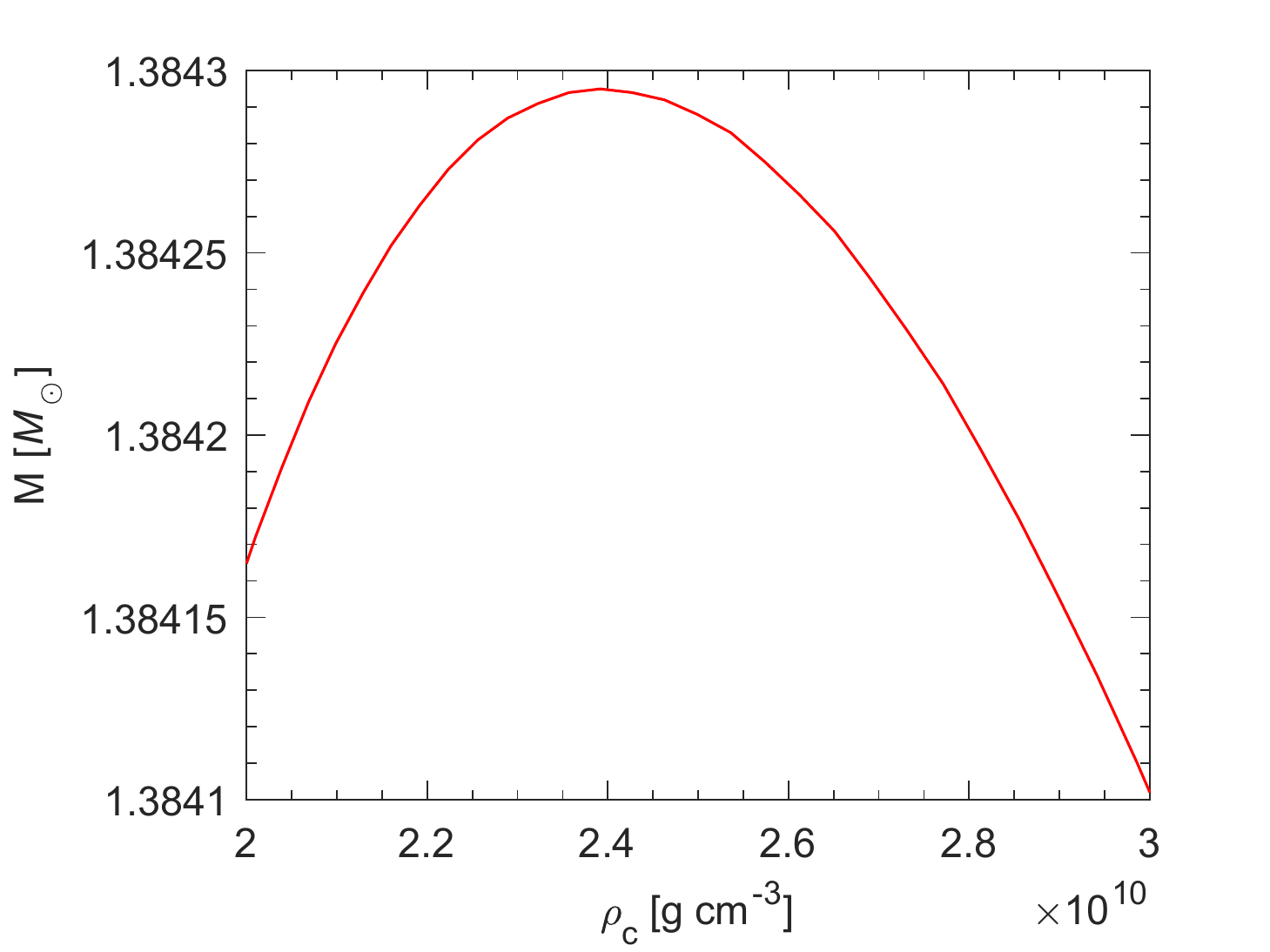}\figsubcap{b}}
		\caption{Gravitational mass in solar masses versus central (mass-energy) density in g~cm$^{-3}$ 
       of carbon white dwarfs in general relativity: (a) comparing results with and without electrostatic corrections, (b) closer view of results with electrostatic corrections around the critical point.}
		\label{fig3}
	\end{center}
\end{figure}

The stability of a white dwarf can be further limited by general relativity, as first shown by Kaplan~\cite{kaplan1949}.
The critical density above which the stellar core becomes unstable can be estimated from the minimization 
of the total energy. Including the electrostatic correction in  Eq.(6.10.28) of Ref.~\citenum{shapiro1983}
we find 
\begin{equation}\label{eq:rhoGR}
\rho_\text{GR} =\frac{16 k_3 (k_2)^2}{(3\pi^2)^{2/3}k_4 (k_1)^2} \frac{M^\prime(A,Z)^2}{Z^2\lambda_e^3 m_e} \biggl[1 + \alpha \frac{8 C_M}{6}\left(\frac{4}{9\pi}\right)^{1/3} Z^{2/3}\biggr]^{-2}~{\rm g~cm}^{-3}\, ,
\end{equation}
where $k_1=1.75579$ and $k_4=0.918294$. 
Because $C_M<0$, electron-ion interactions thus make the star more stable. 
We have also solved the Tolman-Oppenheimer-Volkoff~\cite{tolman1939,oppenheimer1939} equations for calculating the whole sequence of 
white dwarfs in full general relativity. In this way, we have determined more accurately the central (mass-energy) density $\rho_\text{GR}$ at which $dM/d\rho_\text{GR}=0$ marking the onset of instability.  The hydrostatic equilibrium equations were integrated  from the center of the star up to the point where electrons start to bind to nuclei at the density $\rho_\text{eip}=Z M^\prime(A,Z) /a_0^3$ with $a_0$ the Bohr radius. Results 
are collected in Table~\ref{tab5}. The values are systematically lower than those estimated from  Eq.~\eqref{eq:rhoGR}. Comparing Tables~\ref{tab1} and \ref{tab5} shows that the central density is limited  by general relativity rather than electron captures in carbon white dwarfs. Their mass-density relation is plotted in Fig.~\ref{fig3}. The minimum radii and  maximum masses  of white dwarfs with different composition are indicated in Tables~\ref{tab6} and \ref{tab7} respectively 
(note that the mass and radius of carbon white dwarfs at the onset of electron captures are respectively $1.484M_\odot$ and 855~km). 
For comparison, results obtained by solving numerically the Newtonian hydrostatic equilibrium equations are also given. Examining Tables~\ref{tab3} and \ref{tab6} reveals that the polytropic approximation is more reliable for estimating the maximum mass than  the minimum radius.

\begin{table}
	\tbl{Highest density $\rho_\text{GR}$ (in g~cm$^{-3}$) in nonmagnetic stable white dwarfs in general relativity, as calculated by minimization of the approximate total energy (first line) and by solving the hydrostatic equilibrium equations (second line).}
	{\begin{tabular}{@{}cccccc@{}}
			\toprule
			 $^{16}$O              & $^{12}$C             & $^{20}$Ne              & $^{24}$Mg             & $^{56}$Fe   \\
			\colrule
			  $2.73\times 10^{10}$ & $2.72\times 10^{10}$ & $2.75\times 10^{10}$   & $2.76\times 10^{10}$  &  $3.31\times 10^{10}$ \\
			  $2.39\times 10^{10}$ & $2.40\times 10^{10}$ & $2.42\times 10^{10}$   & $2.44\times 10^{10}$  &  $2.93\times 10^{10}$   \\
			\botrule
		\end{tabular}
	}
	\label{tab5}
\end{table}

\begin{table}
	\tbl{Minimum radius (in km) of nonmagnetic stable white dwarfs in Newtonian theory (first line) and general relativity (second line). }
	{\begin{tabular}{@{}cccccc@{}}
			\toprule
			$^{12}$C    & $^{16}$O   & $^{20}$Ne  & $^{24}$Mg & $^{56}$Fe   \\
			\colrule
			$857.7$     & $1055$     & $1447$     & $1736$    & $2082$ \\
			$1010$ & $1052$     & $1444$     & $1732$    & $2079$ \\
			\botrule
		\end{tabular}
	}
	\label{tab6}
\end{table}

\begin{table}
	\tbl{Maximum mass (in solar units) of nonmagnetic stable white dwarfs in Newtonian theory (first line) and general relativity (second line).}
	{\begin{tabular}{@{}cccccc@{}}
			\toprule
			$^{12}$C    & $^{16}$O & $^{20}$Ne & $^{24}$Mg & $^{56}$Fe   \\
			\colrule
		    $1.414$     & $1.403$  & $1.382$   & $1.363$   & $1.118$ \\ 
		    $1.383$  & $1.378$ & $1.366$ & $1.350$  & $1.111$ \\
			\botrule
		\end{tabular}
	}
	\label{tab7}
\end{table}

The influence of a very strong magnetic field on the structure of white dwarfs was 
studied in Ref.~\citenum{das2013} within the theory of polytropes making use of the fact that 
$P\propto \rho^2$ in the strongly quantizing regime. However, it was soon realized  that this  
assumption is unrealistic and that the influence of the magnetic  field itself 
on the stellar structure cannot be ignored~\cite{chamel2013,nityananda2014,coelho2014,caceres2014,bera2014}. Moreover, the 
magnetic-field configuration inside the star should be calculated from 
Maxwell's equations consistently with the stellar equilibrium equations. For all these reasons, 
the structure of magnetic white dwarfs is not easily amenable to a simple analytical treatment.

\begin{figure}[h]%
	\begin{center}
		\parbox{2.4in}{\includegraphics[width=2.7in]{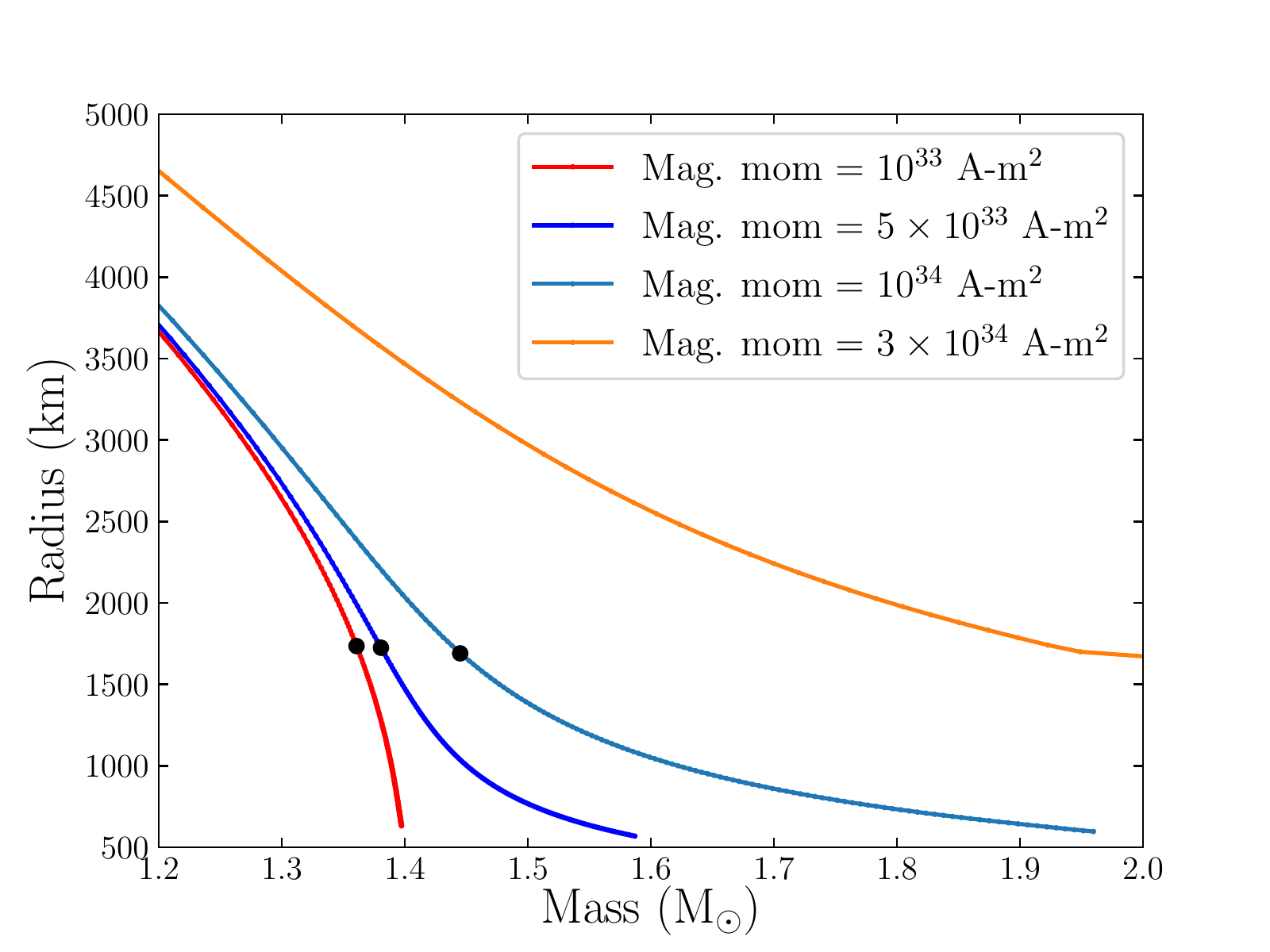}\figsubcap{a}}
		\hspace*{4pt}
		\parbox{2.4in}{\includegraphics[width=2.7in]{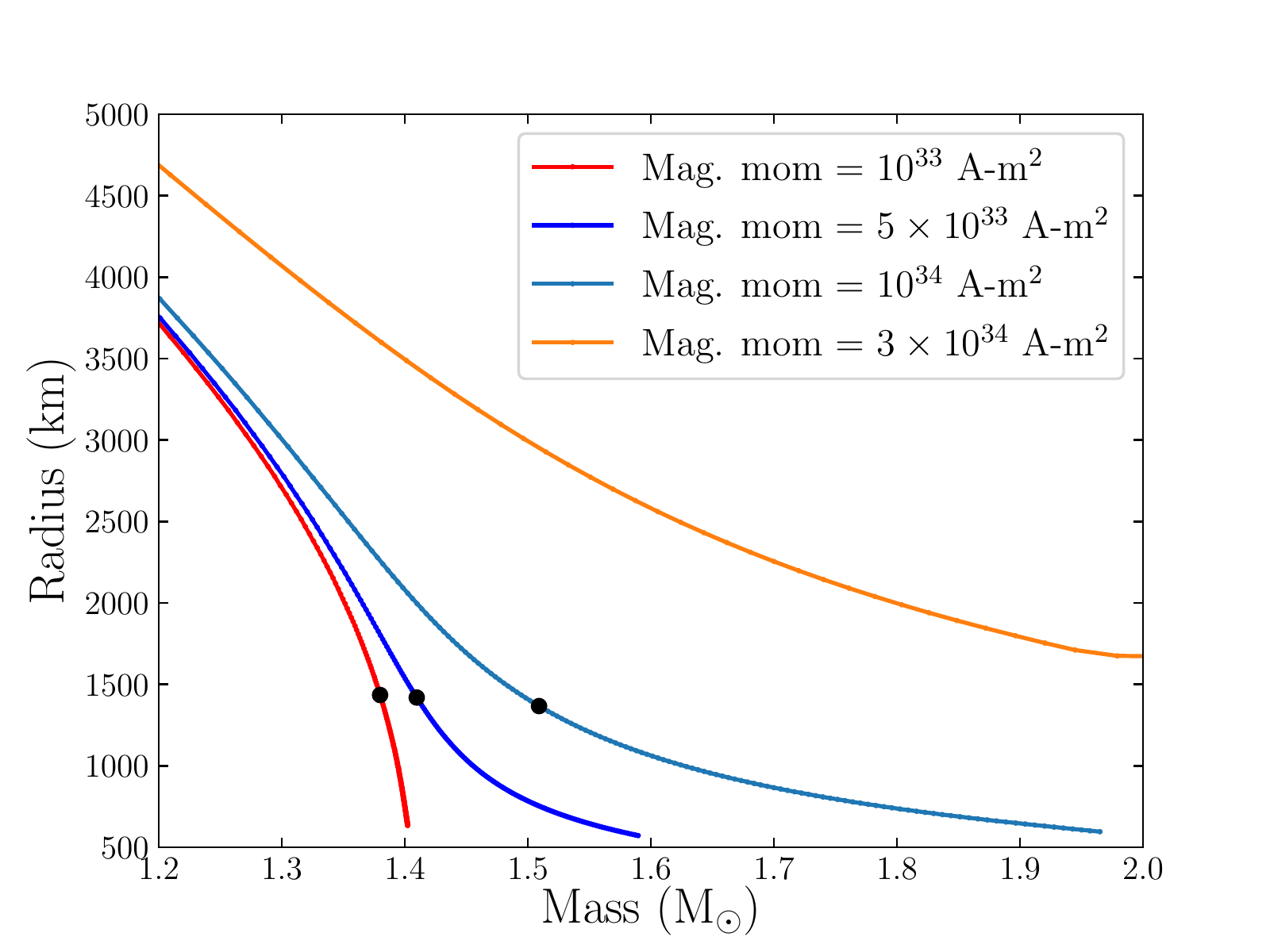}\figsubcap{b}}
		\caption{Radius (in km) versus gravitational mass (in solar units) of white dwarfs for different magnetic moments: (a) for $^{24}$Mg, (b) for $^{20}$Ne. The onset of electron captures is marked by a filled circle.}%
		\label{fig4}
	\end{center}
\end{figure}

To study the influence of strong magnetic fields and the role of electron captures in white dwarfs, we have thus computed fully self-consistently numerical solutions of the Einstein-Maxwell equations using  the \textsc{lorene} library~\cite{gourgoulhon2016}, suitably extended to allow for magnetic-field dependent equations of state and magnetization effects~\cite{chatterjee2015}. We have found that purely poloidal magnetic  fields of order $10^{14}$~G lead to super Chandrasekhar white dwarfs with a mass $\sim 2 M_\odot$. Although such magnetic fields have been found to have a rather small influence on the equation of state, they induce extreme stellar deformations with the most massive white dwarfs adopting a donut-like shape~\cite{chatterjee2017}. Similarly to nonmagnetic white dwarfs, electron captures limit the maximum mass and the minimum radius of magnetic white dwarfs, as summarized in Table~\ref{tab8}. However, the distortion of the star induced by magnetic fields tends to lower their density. For the most extreme configurations, the maximum density thus lies below the threshold density for the onset of electron captures. The mass-radius relations for magnesium and neon white dwarfs are plotted in Fig.~\ref{fig4}. 

\begin{table}
	\tbl{Maximum mass (in solar units) of magnetic white dwarfs, as determined by extreme deformations or electron captures (values in parentheses) for different magnetic moments  $\mathcal{M}$ in A~m$^2$.}
	{\begin{tabular}{@{}cccccccc@{}}
			\toprule
			$\mathcal{M}$      & $^{12}$C            & $^{16}$O          & $^{20}$Ne       & $^{24}$Mg  \\
			\colrule
			$10^{33}$          & $1.41$ ($1.41$)     &  $1.41$ ($1.40$)  & $1.40$ ($1.38$) & $1.39$ ($1.38$) \\ 
			$5\times 10^{33}$  & $1.60$ ($1.50$)     &  $1.59$ ($1.46$)  & $1.59$ ($1.41$) & $1.58$ ($1.36$) \\ 
			$10^{34}$          & $2.00$ ($1.86$)     &  $1.97$ ($1.67$)  & $1.97$ ($1.51$) & $1.96$ ($1.45$) \\ 
			$2\times 10^{34}$  & $1.99$ ($1.99$)     &  $1.96$ ($1.96$)  & $1.97$ ($1.95$) & $1.95$ ($1.73$)  \\ 
			$3\times 10^{34}$  & $1.96$ ($1.96$)     &  $1.94$ ($1.94$)  & $1.98$ ($1.98$) & $1.97$ ($1.97$)  \\
			\botrule
		\end{tabular}
	}
	\label{tab8}
\end{table}

\begin{figure}[h]
	\begin{center}
		\includegraphics[width=2.7in]{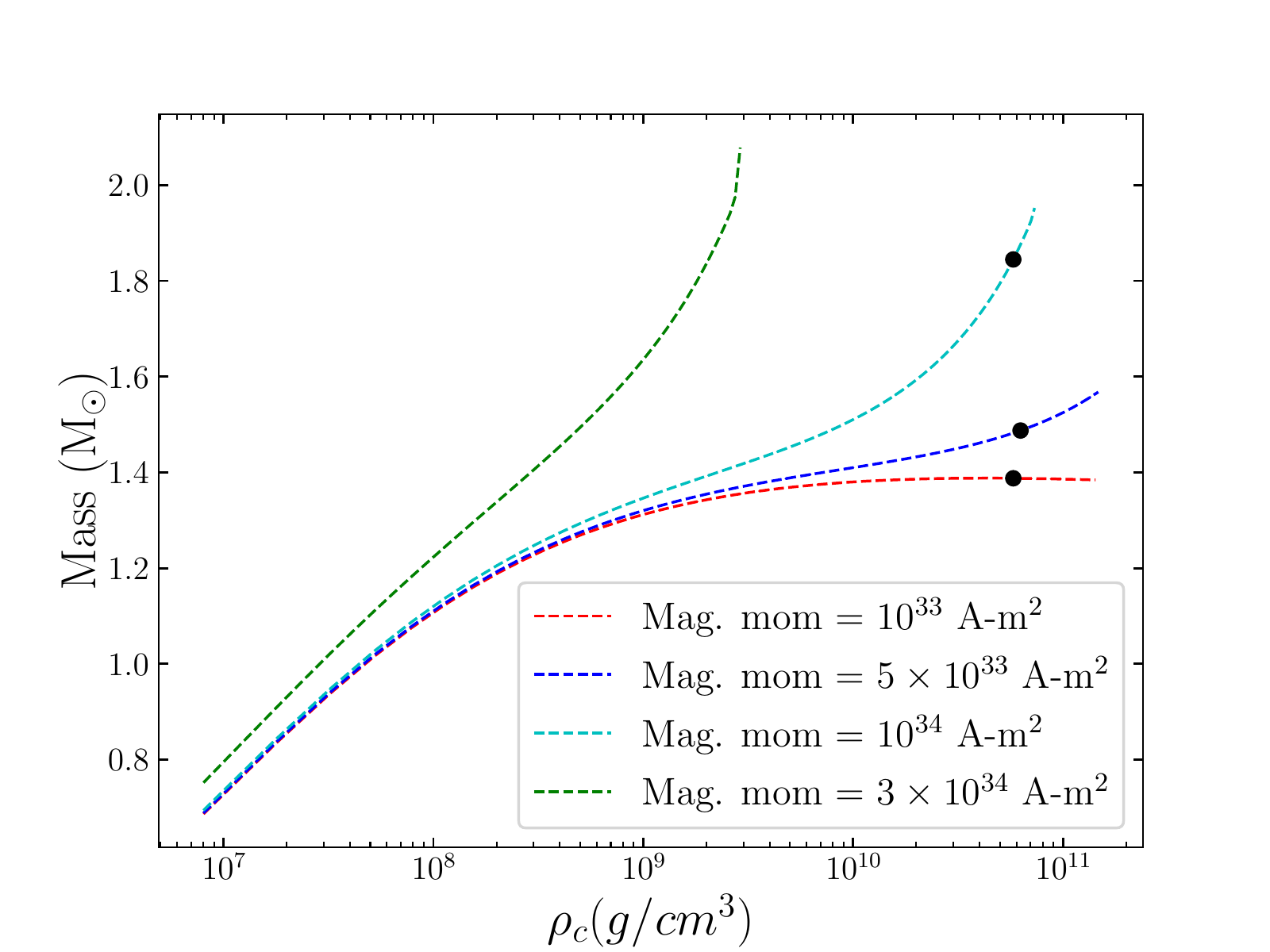}
	\end{center}
	\caption{Gravitational mass (in solar masses) versus central (mass-energy) density in g~cm$^{-3}$ 
		of carbon white dwarfs in general relativity with given magnetic moments. The onset of electron captures is marked by a filled circle.}
	\label{fig5}
\end{figure}

The global stability criterion $\partial M/\partial \rho>0$ still remains valid for magnetic white dwarfs provided the derivatives are evaluated for a fixed magnetic moment $\mathcal{M}$. As shown in Fig.~\ref{fig5}, this instability is entirely removed by the presence of strong magnetic fields. In other words, the magnetic field makes the star more stable. In particular, the maximum mass of magnetic white dwarfs made of carbon is not limited by general relativity as their nonmagnetic relatives but by electron captures. 

For magnetic white dwarfs to be the progenitors of overluminous type Ia supernova, i.e. to have masses $\sim 2M_\odot$, their magnetic field must be strong enough. Their minimum observable polar magnetic fields are indicated in Table~\ref{tab9} for different compositions. In all cases, the magnetic dipole moment is $3\times 10^{34}$~A~m$^2$.

\begin{table}
	\tbl{Polar magnetic field (in $10^{13}$~G) of white dwarfs with $\mathcal{M}=3\times 10^{34}$~A~m$^2$.}
	{\begin{tabular}{@{}ccccccc@{}}
			\toprule
			$^{12}$C  & $^{16}$O & $^{20}$Ne  & $^{24}$Mg  \\
			\colrule
			$2.73$    &  $2.55$  & $2.89$     & $2.85$   \\ 
			\botrule
		\end{tabular}
	}
	\label{tab9}
\end{table}

\section{Conclusions} 

The stability of white dwarfs is limited by  the onset of electron captures by  nuclei in their core. 
We have presented  very accurate analytical formulas for calculating the threshold electron Fermi energy, mass  density and pressure 
in cold dense Coulomb plasmas with magnetic fields, taking into 
account Landau-Rabi quantization of electron motion. In particular, the density exhibits typical quantum oscillations associated with the filling of energy levels. As a consequence, electron captures in  magnetic white dwarfs may occur at a lower or higher  density than in their nonmagnetic relatives depending on the magnetic field strength. The lowest possible density is found to lie about 25\% below its value in  the absence of magnetic field essentially independently  of the  composition.  On the contrary, the density is not limited from above and increases  almost linearly  with the magnetic field strength in the strongly quantizing regime. 

Taking into account electron-ion interactions using the polytropic approximation $P\approx K\rho^{4/3}$ with $K$ given by Eq.~\eqref{eq:K},  we have explicitly shown how electron captures alter the maximum mass of nonmagnetic white dwarfs and set a lower limit to  their radius. We  have also solved numerically  the hydrostatic equilibrium equations using the full equation of state, both in Newtonian theory and in general relativity. We have found that electron captures reduce the maximum mass of white  dwarfs by 3-13\%  compared to the Chandrasekhar model. 

Solving the Einstein-Maxwell equations taking into account the magnetization of dense matter, we  have found that white dwarfs with purely poloidal magnetic fields can be significantly more massive than their nonmagnetic relatives. We have also shown that the presence of strong magnetic fields makes the star more stable. The maximum mass of magnetic white dwarfs (including those made of carbon) is thus solely limited by electron captures. In turn, the large stellar deformations induced by the strongest magnetic fields lower the stellar density below the electron capture threshold. White dwarfs with polar magnetic fields exceeding $10^{13}$~G could thus be massive enough to explain overluminuous type Ia supernova independently of the composition of their core. On the other hand, a purely poloidal magnetic-field configuration is unstable. The magnetic field in super Chandrasekhar white dwarfs is expected to have both poloidal and toroidal components. The question as to whether such stars can exist still remains open. 

Although the present study was focused on white dwarfs, it may be also of interest for neutron stars as electron captures by nuclei constituting the outer crust of magnetars may explain their persistent luminosity and outbursts~\cite{chamel2021}. The general analytical formulas for the threshold density and pressure could thus be applied to determine the precise locations of these reactions.

\section*{Acknowledgments}
The work of N.C. was financially supported by F.R.S.-FNRS under Grant No. IISN 4.4502.19. L. P. is a FRIA grantee of F.R.S.-FNRS. This work was also supported by COST CA16214 and the CNRS International Research Project ``Origine des \'el\'ements lourds dans l’univers: Astres Compacts et Nucl\'eosynth\`ese (ACNu)''.

\bibliographystyle{ws-procs961x669}
\bibliography{references}

\end{document}